\documentclass[amsmath,letter]{aa}

\usepackage{amsmath}
\usepackage{txfonts}
\usepackage{natbib}
\usepackage{graphicx}

\newcommand{\bdes}{\begin{description}}
\newcommand{\edes}{\end{description}}

\begin{document}

\title{Cold CO in circumstellar disks}
\subtitle{On the effects of photodesorption and vertical mixing}

\author{F. Hersant\inst{1,2}, V. Wakelam\inst{1,2}, A. Dutrey\inst{1,2}, S. Guilloteau\inst{1,2} \and E.
Herbst\inst{3} }

\offprints{Franck Hersant \email{Franck.Hersant@obs.u-bordeaux1.fr}}

\institute{
Universit\'e de Bordeaux, Laboratoire d'Astrophysique de Bordeaux (LAB)
\and
CNRS/INSU - UMR5804 ; BP 89, F-33271 Floirac Cedex, France
\and
Departments of Physics, Astronomy, and Chemistry, The Ohio State University, Columbus, OH 43210, USA
}

\date{Received ??? / Accepted ???}

\abstract
{}
{We attempt to understand the presence of gas phase CO below its sublimation temperature in circumstellar disks.
We study two promising mechanisms to explain this phenomenon: turbulent mixing and photodesorption.}
{We compute the chemical evolution of circumstellar disks including grain surface reactions with and without
turbulent mixing and CO photodesorption.}
{We show that photodesorption significantly enhances the gas phase CO abundance, by
extracting CO from the grains when the visual
extinction remains below about 5 magnitudes. However, the resulting
dependence of column density on radial distance is inconsistent with observations so far. We propose that this inconsistency could be
the result of grain growth. On the other hand, the influence of turbulent
mixing is not found to be straightforward. The efficiency of turbulent mixing depends upon a variety of parameters, including  the
disk structure. For the set of parameters we chose, turbulent mixing is not found to
have any significant influence on the CO column density.}
{}


\maketitle

\section{Introduction}

In circumstellar disks, because of the high densities, carbon monoxide is expected to stick onto grains at
temperatures below about 17 K. However,
millimeter interferometric observations of several circumstellar disks around T Tauri stars (e.g. DM
Tau, GM Aur, and LkCa15) have revealed that a large amount of CO remains gaseous at temperatures as low
as 10 K \citep{Dartois03,Pietu07}. These authors show that the bulk of CO is at temperatures
below 17 K between 100 and 800 AU from the central star.
Since these disks are much older than the sticking timescale of
CO onto grains ($\sim 10^2$ yr), some mechanism must prevent the complete depletion of CO. 

\cite{Aikawa06} proposed that the presence of cold CO in disks can be a consequence of vertical turbulent
mixing \citep[See also Fig. 4 of][]{Willacy06}. \cite{Semenov06} studied in detail the
influence of both vertical and radial mixing on the abundance of cold CO. They found that vertical mixing cannot
account for the observed abundance of cold CO, while the situation improves slightly when considering
the full 2D mixing (vertical and radial).
However, \cite{Aikawa07} computed analytically the influence
of vertical mixing on the abundance of cold CO and concluded that vertical mixing was sufficient to achieve an
abundance of CO consistent with that observed by \cite{Dartois03}. 
The apparent discrepancy with \cite{Semenov06} can be understood as a consequence of the different efficiencies of
grain surface reactions in their studies \citep{Aikawa07}.

On the other hand, \cite{Oberg07} measured  the photodesorption rate of CO in the laboratory. They found 
a rate similar to that found for H$_2$O by \cite{Westley95},
about 2 orders of magnitude higher than previously available theoretical estimates \citep{Draine79}, implying
that this non-thermal
desorption process was an appealing candidate for explaining the abundance of cold CO in various astrophysical environments.
This motivated us to investigate the influence of both vertical mixing and photodesorption on the CO column
density in disks, using a parametric model consistent with interferometric observations of circumstellar
disks.

In Sect. \ref{model}, we describe the physical and chemical model used, while Sect.
\ref{results} is devoted to the results of our computations. In Sects. \ref{discussion} and
\ref{conclusions}, we discuss our results and present our conclusions.

\section{Modeling}
\label{model}

\subsection{Disk structure}

To consistently compare with the physical structure observed by \cite{Pietu07},
we use a simple disk model with physical quantities varying as power laws of the radius, similar to the
parametric disk model used by \cite{Pietu07} to reproduce their interferometric maps. The vertical
temperature distribution is defined to have two layers, one cold layer with a constant temperature of 10 K, and
a warm layer in radiative equilibrium with the central star radiation. 
This behavior is typical of disk models \citep[e.g.][]{Alessio99,
Aikawa06}. The precise location of the transition between the two layers varies between models.
We chose this transition 
at 2 pressure scale heights, a location similar to that used by \cite{Dartois03} to fit their data.
The constant mid-plane temperature,
although surprising at first sight, is
consistent both with the temperature observed in $^{13}$CO J=1-0 by \cite{Pietu07} and the surface
density power exponent of $-3/2$ observed in the outer parts of most ``undisturbed" disks \cite[see][for a
review]{Dutrey08}. In an $\alpha$-model \citep{SS73}, the product of the temperature and the surface density
is indeed assumed to vary as R$^{-3/2}$. Therefore, a constant mid-plane temperature in an stationary $\alpha$-model is the only way 
that the surface density decreases as R$^{-3/2}$. This regions of almost constant temperature
are also found in the more complete disk models developed by \cite{Alessio99, Alessio01}.

The laws that we use in the present paper are:
\begin{eqnarray}
T(R,z)&=&T_C + (T_H(R)-T_C)\times \frac{1}{2}\left(1+tanh\left(\frac{z - 2 H(R)}{H(R)/3}\right)\right)\\
T_C&=&10\ \mathrm{K}\\
T_H(R)&=& 30\ \left(\frac{R}{100\ AU}\right)^{-1/2}\ \mathrm{K}\\
H(R)&=& \frac{c_s(T_C)}{\Omega} \simeq 8.4 \left(\frac{T_C}{10\ K}\right)^{1/2}\left(\frac{R}{100\
AU}\right)^{3/2} \mathrm{AU}\\
\Sigma(R)&=&0.8\ \left(\frac{R}{100\ AU}\right)^{-3/2} \mathrm{g\ cm^{-2}}
\end{eqnarray}
where $c_s$ is the isothermal sound speed, $\Omega$ is the Keplerian frequency, and $\Sigma$ is the surface
density. From these laws, the mass density $\rho$ is then computed from the hydrostatic equilibrium.


\subsection{Chemistry}

The chemical evolution of the disk is computed with a 1D vertical model that takes into account both gas-phase
and grain surface chemistries. This code was adapted from the OSU gas-grain code developed by the
astrochemistry team of Eric Herbst. The formalism of surface  chemistry and various processes are described in
\cite{Hasegawa92}  and \cite{Garrod07}. The gas-phase chemical network uses updates from
\verb+osu_03_2008+\footnote{http://www.physics.ohio-state.edu/\~{}eric/research.html}. The complete network
(gas and grain) contains 655  species and 6067 reactions. Grains are assumed to be of interstellar type. To
model the CO and H$_2$ UV self-shielding, we use
the approximation from \cite{Lee96}. The code is written in Fortran 90 and uses the DLSODES routine from
the ODEPACK package \citep{Hindmarsh83} to solve the differential equations. The initial abundances of our
calculations are taken from the output of a gas-phase calculation of a dark cloud ($n_{\rm H}$=$2\ 10^4$
cm$^{-3}$, $T=10$ K, $A_{\rm V}$= 10) after $10^7$ yr and our elemental abundances are the EA3 conditions from
\cite{Wakelam08}, with an S abundance of $8\times 10^{-8}$.  The gas-grain model is then allowed to evolve for $10^5$ yr
with 32 vertical points. A period of $10^5$ yr is the radial mixing timescale at 100 AU, as well as the
evolutionary timescale for the disk structure. Computing for a longer time would in principle require a full
2D (R,z) approach and an evolving disk model.

\subsection{Photodesorption}
In their experiments, \cite{Oberg07} found a photodesorption rate Y$_{PD}$ of $3\times 10^{-3}$ CO molecules
per UV photon above pure CO ice. In circumstellar disks, UV photons originate from two main processes: (i)
the ambient UV field, coming from both the central star and the surrounding ISM, and (ii) secondary photons
generated by the interaction of H$_2$ with cosmic rays and cosmic ray-induced (CR) secondary electrons
\citep{Prasad83,Shen04}. Photodesorption of CO (we did not consider photodesorption for other species) by these two sources of UV photons can be expressed by the chemical
reactions: 
\begin{eqnarray} 
&\mathrm{adsorbed\ CO} \xrightarrow[]{k_{UV}} \mathrm{CO}, \\ 
&\mathrm{adsorbed\ CO} \xrightarrow[]{k_{CR}} \mathrm{CO}, 
\end{eqnarray} 
where $k_{UV}$ and $k_{CR}$ are the first-order rate
coefficients (s$^{-1}$). From \cite{Oberg07} and the additional approximation that adsorbed CO molecules can
always escape independently of the thickness of the adsorbed layer, we can express these rate coefficients as:
\begin{eqnarray} 
&k_{UV}=\frac{Y_{PD}}{\sigma_{sites}}\ I_{ISM-FUV}\ f_{UV}\ e^{-\gamma\ A_V}\\
&k_{CR}=\frac{Y_{PD}}{\sigma_{sites}}\ I_{CR-FUV} 
\end{eqnarray} 
(Hassel et al. in preparation) where
$I_{ISM-FUV} \simeq 10^{8}$ photons cm$^{-2}$ s$^{-1}$ is the ambient far UV field strength for a standard ISM
irradiation field \citep{Draine78}, $I_{CR-FUV} \simeq 10^{4}$ photons cm$^{-2}$ s$^{-1}$ is the strength of
the far UV field generated by the cosmic rays, $\gamma$ measures the difference between the UV extinction and
the visual extinction \citep[$\gamma \sim 2$ for interstellar grains,][]{Roberge91}, and $\sigma_{sites}$ is
the site density per grain surface unit (i.e. the maximum number of molecules adsorbed as a single layer on a
grain per cm$^{2}$). The quantity $f_{UV}$ is a scale factor used to express the strength of the ambient UV
field (originating mainly in the central star) as a multiple of the standard ISM radiation field. This expression
is justified by noting that the UV spectrum coming from the UV excess of young stars has a similar shape to
that of the standard interstellar field \citep{Chapillon08}. In the present report, we use: \begin{equation}
f_{UV}=300\ \left(\frac{\sqrt{R^2+(4 H)^2}}{100\ AU}\right)^{-2} \end{equation}

The UV field intensity is consistent with the results of \cite{Bergin03}.
We consider here only the vertical extinction of the UV field. This approximation is motivated by the fact
that the direct (radial) extinction of stellar UV radiation is very efficient so that most photons seen by the disk are in
fact scattered by small grains at high altitudes. We consider here that this scattering acts at the edge of our
computation box (4 pressure scale heights).
With this law for $f_{UV}$, direct photodesorption is dominant for Av $\lesssim$ 7,
while photodesorption by CR-induced UV photons (independent of Av) maintains a low desorption rate in
denser regions.

\subsection{Turbulent mixing}

We compute the vertical turbulent mixing using an $\alpha$ turbulent viscosity \citep{SS73}, where the
diffusivity $D_t$ is given by:
\begin{equation}
D_t= \alpha\ \mathrm{Sc}\ \frac{c_s^2}{\Omega}
\end{equation}
The Schmidt number Sc,
defined as the ratio of the turbulent viscosity to the turbulent diffusivity, is a poorly constrained
parameter. In the present work, we set its value to be $1$ and choose $\alpha=10^{-2}$, values commonly used
(e.g. \citet[][]{Semenov06,Willacy06,Aikawa07}).\\

The resulting equations for the chemical abundances $x_i$ are:
\begin{equation}
\frac{\partial x_i}{\partial t} = P_i - L_i +\frac{1}{\rho} \frac{\partial}{\partial z} D_t
\rho\frac{\partial}{\partial z} x_i
\end{equation}
where $P_i$ and $L_i$ are the chemical production
and loss terms.\\
The vertical mixing and the chemistry are computed using first-order operator splitting, the chemistry being
computed  before
the mixing inside a time step. This induces intrinsic errors, which are controlled by reducing the time step. We
adapted the time step so that the splitting errors remain below 5 \% relative accuracy, by comparing with
simulations without operator splitting, which are far more time-consuming.

\section{Results}
\label{results}

\begin{figure}
\centering
\includegraphics[angle=270,width=0.45\textwidth]{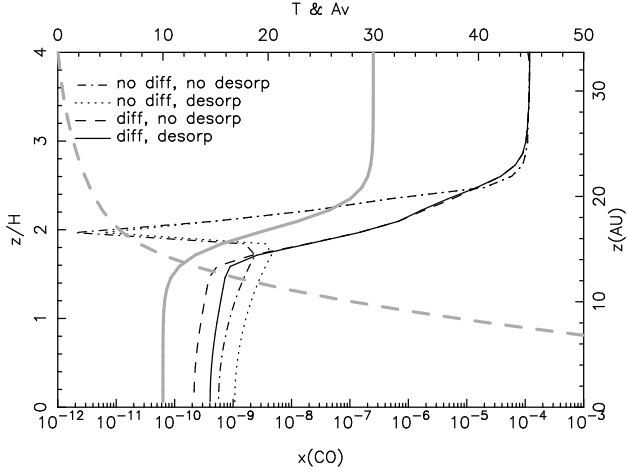}
\caption[]{Vertical distribution of gaseous CO at 100 AU from the central star, 
(a) no photodesorption, no diffusion (dotted-dashed line),
(b) no photodesorption, diffusion (dashed line),
(c) photodesorption, no diffusion (dotted line),
(d) photodesorption, diffusion (solid line). Superimposed on the graph are the temperature (thick
grey line) and visual extinction (thick dashed grey line) as a function of height above
the midplane.}
\label{co100}
\end{figure}

\begin{figure}
\centering
\includegraphics[angle=270,width=0.45\textwidth]{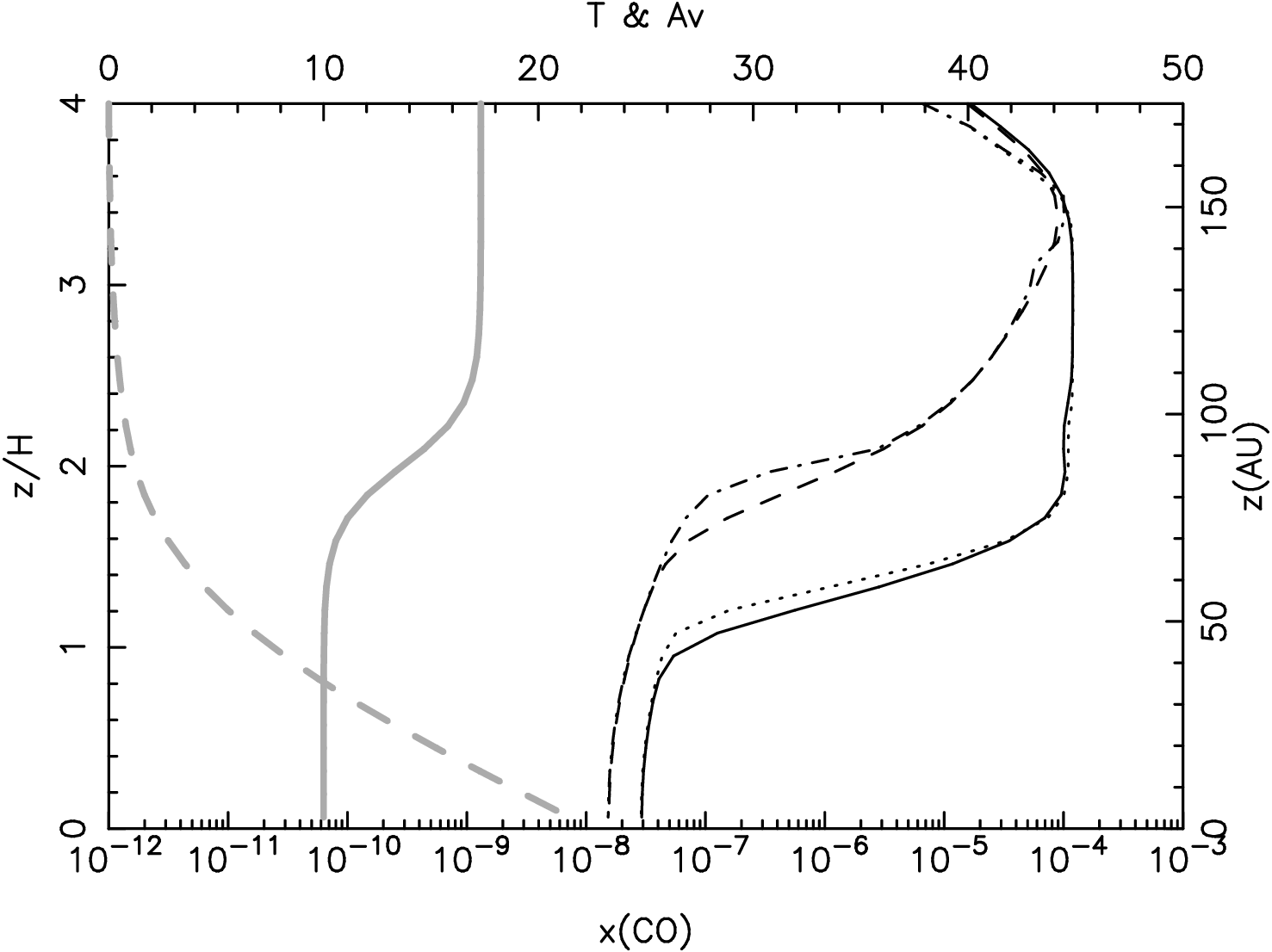}
\caption{As Fig. 1 but for R=300 AU}
\label{co300}
\end{figure}

\begin{figure}
\centering
\includegraphics[angle=270,width=0.45\textwidth]{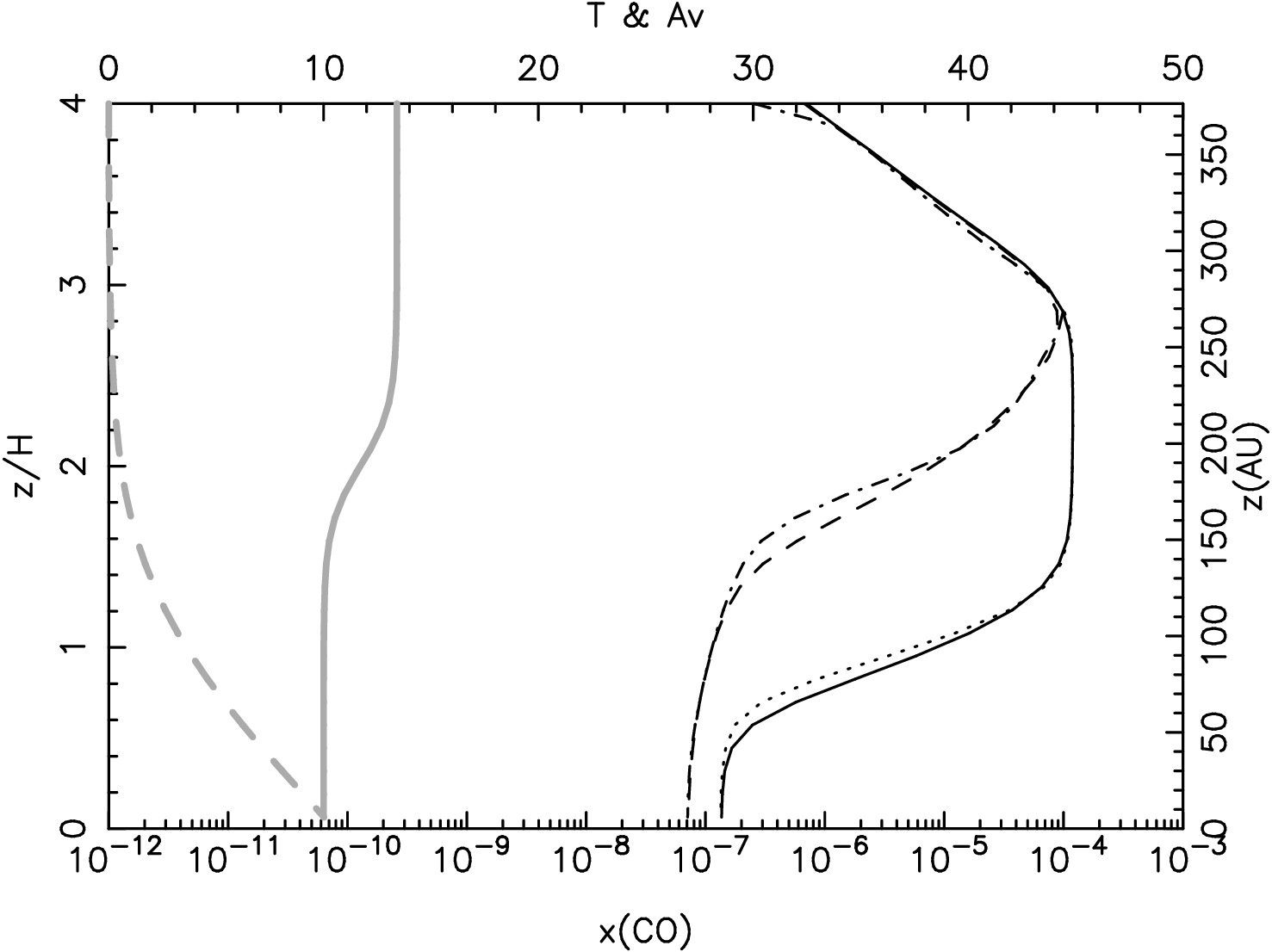}
\caption{As Fig. 1 but for R=500 AU}
\label{co500}
\end{figure}

The vertical distribution of gas phase CO was computed at 100, 200, 300, 400, 500, and 600 AU, for 4 different
models, with or without CO photodesorption and/or vertical mixing. Figures \ref{co100}, \ref{co300}, and
\ref{co500} show the resulting vertical CO distributions. The ``handle-bar" shape of the vertical
abundance profile at all radii is the result of CO photodissociation at high altitudes (mostly visible at 300
and 500 AU) and CO freezing onto grains in the dense mid-plane (below $z/H \simeq 2$). Carbon is locked
in different dominant forms as a function of height:
frozen CO in
the midplane, then frozen CO$_2$, gaseous CO, and finally atomic C in the photodissociated region. The 
thickness of each layer depends on the radius (through density, temperature, and Av). In the 100 AU profile (Fig.
\ref{co100}), a dip is visible in the CO distribution without mixing. This dip is the consequence of a
layer of N$_2$H$^+$, which initiates a secondary reaction chain (dominated by grain chemistry) started by: 
\begin{equation}
\mathrm{N_2H^+ + CO} \xrightarrow[]{} \mathrm{N_2 + HCO^+}.  
\label{n2h+} 
\end{equation} 

\begin{figure}
\centering
\includegraphics[angle=270,width=0.45\textwidth]{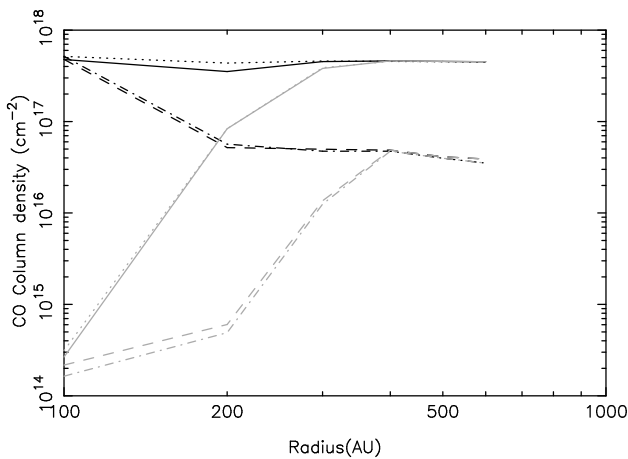}
\caption[]{Total column density of gaseous CO as a function of the radial distance,
(a) no photodesorption, no diffusion (dotted dashed line),
(b) no photodesorption, diffusion (dashed line),
(c) photodesorption, no diffusion (dotted line),
(d) photodesorption, diffusion (solid line). The grey lines are the corresponding cold CO (T $<$ 17 K) column
density profiles.}
\label{cosigma}
\end{figure}

HCO$^+$ is then converted into other carbon-bearing molecules, the reaction chain ending with frozen
CO$_2$.
The  CO column
density as a function of radius is displayed in Fig. \ref{cosigma}.

\subsection{Influence of photodesorption}

Photodesorption by primary UV photons is the most important process affecting the abundance of gas-phase CO. When the visual
extinction is sufficiently low (below about 5 mag), this desorption mechanism is efficient enough to preserve a
significant abundance of CO. For radii larger than 200 AU,  the total column density of gas phase CO is
increased by an order of magnitude when this process is taken into account. The situation is different at
100 AU. Here, the visual extinction reaches a value of 5 mag at 2 pressure scale heights, and
grains receive insufficient primary UV photons to enable efficient photodesorption of CO. The secondary UV photons
originating from the interaction of H$_2$ with electrons induced by cosmic rays are then the only
photodesorbing photons. They maintain a small fraction of CO in the gas phase but do not influence the total
CO column density by more than 10 \%.

\subsection{Influence of vertical mixing}

In our simulations, vertical mixing has a small influence on the total column densities of CO. At 100 AU
(and 200 AU), as shown in Fig. \ref{cosigma}, the presence of a layer of N$_2$H$^+$ appears as a sink of CO
(via reaction \ref{n2h+}) and the total CO column density is reduced by vertical mixing.
This is the reason why the CO abundance below $z=2H$ is lower in the case with diffusion (Fig. \ref{co100}). 
The low efficiency of vertical mixing is a consequence of two effects. First, when CO condenses onto
grains, grain chemistry transforms it into more refractory species, thereby preventing desorption when it
diffuses back to the warm region.
The second reason is the
consequence of a dilution effect. Due to a temperature transition at 2 pressure scale heights, the warm region
represents less than 10 \% of the total mass. This is obviously a possible bias of our disk model, the influence of
vertical mixing being very sensitive to this transition. For a transition deeper inside in the disk, as
considered by \cite{Aikawa06} and \cite{Aikawa07}, vertical mixing is far more efficient. Accurate knowledge
of the true thermal structure of the disk would be required to obtain a more precise estimate.

\section{Discussion}
\label{discussion}




Figure \ref{cosigma} shows that photodesorption increases the total CO column density, but also makes the CO
column density flatter. \cite{Pietu07} measured a steep radial profile of CO column density, varying as
R$^{-3}$ or even more steeply, with about $10^{17}$ cm$^{-2}$ at 300 AU. This gradient is steeper than for our models without photodesorption.
Quantitative comparisons with observations are however far from being straightforward because they  require
the generation of CO emission maps and iterations on the physical structure to find the best fit for each
chemical setup. Incidentally, the flat CO column density profile is a common feature of disk chemical
models. In our case, we would like to raise a few points:

\bdes
\item[-] Our results are extremely sensitive to the location of the transition between the cold and warm
regions. We chose a transition at 2 pressure scale heights; other choices would lead to
significant differences in the column density radial profile.
\item[-] At large distances from the central star, the CO column density is very sensitive to 
photodissociation. The stellar UV flux is not constrained accurately, although its value can  significantly
alter our results.
Our treatment of the UV field is also approximate. We only consider UV radiation in a direction perpendicular to
the disk. Observations of $^{12}$CO are sensitive only to the outer regions \citep[][]{Pietu07}, where the interstellar UV
field impinging from the disk edge will steepen the slope of the CO column density profile.
\item[-] For a given UV flux, photodissociation is controlled by the relation between visual extinction and
column density. This factor depends mostly upon the amount of small grains. Grain growth can change the penetration
of UV radiation inside the disk and dissociate CO further \citep[see][]{Chapillon08}. Since grain growth rates depend upon density, the grain
distribution can depend upon the radial distance and completely change the radial profile of the CO column
density. In turn, it may be possible to invert the CO column density to infer some information about the current state of grain
growth at a given radius.  
\item[-] Grain growth also increases the timescales for the adsorption of CO and grain surface chemical
reactions, making both photodesorption and vertical mixing more efficient at
retaining CO in the gas phase.
\edes

\section{Conclusions}
\label{conclusions}

We have shown that CO photodesorption appears to be a good candidate to explain the
abundance of cold gas phase CO in circumstellar disks. However, it tends to create a flatter radial profile for the CO
column density, in contrast to that observed in disks. Therefore, more work is necessary to solve this problem.
Consideration of the photodissociating UV flux and variations in its penetration inside the disk due to grain
growth and geometry seem to provide an avenue to improve our modelling of circumstellar disks. Moreover, we have shown
that the influence of vertical mixing on the abundance of CO is not straightforward.  Rather, the
efficiency of vertical mixing depends upon many parameters, 
including the details of the disk structure (density and temperature) and the level of grain growth, all of
which remain uncertain today.

\begin{acknowledgements}
FH was supported by a CNRS fellowship. FH wishes to thank Alan Hindmarsh for useful advice on the use of
LSODES. We thank the referee Yuri Aikawa for constructive remarks which helped us to improve the clarity
of the paper. FH, VW, AD and SG are financially supported by the French program "Physique Chimie du Milieu
Interstellaire" (PCMI). EH thanks the National Science Foundation (US) for its support of his program in
astrochemistry.
\end{acknowledgements}

\bibliographystyle{aa}
\bibliography{cobib}

\end{document}